\documentclass{aa}

\usepackage{graphicx}
\graphicspath{{./}{figures/}}
\usepackage{natbib}
\usepackage{amsmath}
\usepackage{xcolor}
\usepackage{txfonts}
\usepackage[colorlinks=true]{hyperref}
\hypersetup{linkcolor=blue,citecolor=blue,filecolor=blue,urlcolor=blue}
\begin{document} 

\title{Radiation pressure on dust explains the Low Ionized Broad Emission Lines in Active Galactic Nuclei}

\subtitle{Dust an important driver of line shape}

   \author{M. H. Naddaf
          \inst{1,2}\fnmsep\thanks{naddaf@cft.edu.pl}
          \and
          B. Czerny
          \inst{1}\fnmsep\thanks{bcz@cft.edu.pl}
          }

   \institute{Center for Theoretical Physics, Polish Academy of Sciences, Lotnik\'ow 32/46, 02-668 Warsaw, Poland\\
              %\email{}
         \and
             Nicolaus Copernicus Astronomical Center, Polish Academy of Sciences, Bartycka 18, 00-716 Warsaw, Poland\\
             %\email{}
             %\thanks{}
             }

\date{Accepted: April 11, 2022}

% \abstract{}{}{}{}{} 
% 5 {} token are mandatory
 
  \abstract
   {Broad emission lines are the most characteristic features in the spectra of galaxies with active nucleus (AGN). They mostly show either single-peaked or double-peaked profiles; and originate from a complex dynamics of the likely discrete clouds moving in a spatially extended region so-called Broad Line Region (BLR).}
   {In this paper, we provide a large grid of results based on which we aim at testing the model with calculation of the spectral line generic profiles.}
   {We follow a non-hydrodynamical single-cloud approach to the BLR dynamics based on the radiatively dust-driving model of Czerny \& Hryniewicz. We previously showed in detail that the 2.5D version of the model could provide us with the 3D geometry of the BLR.}
   {We show that the shape of profiles not only depends on the accretion rate of the source, the black hole mass, and the viewing angle, but also it is most significantly affected by the adopted dust-to-gas mass ratio regulating the strength of the radiation pressure. We also show that the model can appropriately explain the low ionized broad lines of the mean spectrum of quasars, such as MgII and H$\beta$.}
   {The radiatively dust-driving mechanism can appropriately account for the low-ionized part of BLR of AGNs.}


\titlerunning\space
\maketitle

%------------------------------------------------------

\section{Introduction}

Broad Line Region (BLR) in active galaxies is a turbulent and spatially rather extended region \citep{wandel1999,kaspi2000,netzer2020} which has not yet been fully resolved, except of a few cases with magnified BLR due to gravitational lensing \citep{sluse2012, guerras2013} and the VLR GRAVITY observations in the IR \citep{GRAVITY3C273_2018, GRAVITYIRAS_2020, GRAVITY3783_2021}. Hence, studies of broad emission lines stemming from this region have been based on the analysis of the spectra, and their time dependence \citep{Boroson1992, Lawrence1997, Sulentic2000, Reeves2000, Gaskell2009, Le2019, Raimundo2020}. However, the broad emission lines (BELs) from BLR are a unique probe to understand the physics of AGNs and eventually can help to measure the mass of the central supermassive black hole. These emission lines are divided into two categories of High Ionization Lines (HIL) and Low Ionization Lines (LIL) \citep{CollinSouffrin1988, netzer2013}; and they are frequently observed in the form of single-peaked or double-peaked profiles \citep{osterbrock1977, osterbrock1981, gezari2007, shen2011, negrete2018, zhang2019, lu2021, li2021}. The shape of BELs depends on the distribution and dynamics of material within BLR. There have been two mainstream studies for years on the theoretical side to recover the observed shape of emission lines: one based on proposing theoretically-motivated mechanisms for the formation of BLR, and the other one on the basis of assuming a certain distribution of material in BLR, so-called parametric models.

\begin{figure*}
	\centering
	\includegraphics[scale=0.46]{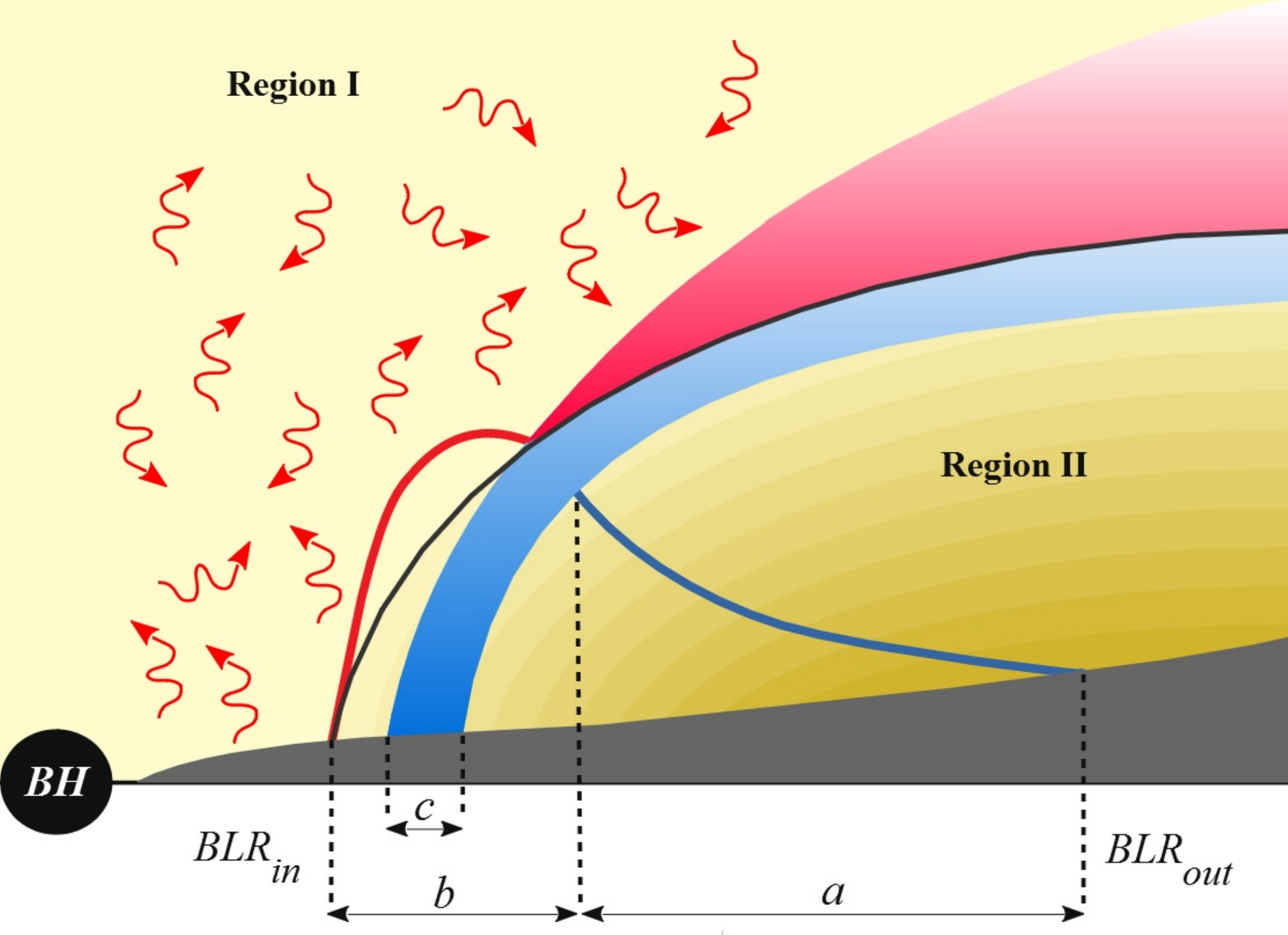}
	\caption{The schematic illustration of BLR shape and its components in FRADO. The components of BLR are represented as (a) an outer tail consisting of dusty failed winds (simple up/down motions), (b) an inner region with slightly to largely elongated elliptical orbits, and (c) a funnel-shaped outflow stream of clouds. The red and blue solid lines show the peak height of the trajectories of dustless and dusty clouds, respectively. The $BLR_{\rm in}$ and $BLR_{\rm out}$ set the two ends of a radial region of the accretion disk within which the material can be lifted due to disk radiation pressure. The black solid line represents the sublimation location. The {\it Region I} marks the hot region above the sublimation location where dust can not survive; the central disk UV radiation reaches the clouds via scattering by the hot gas in this region. The {\it Region II} underneath of the sublimation location and cold enough for dust survival is obscured by the outflow stream; it prevents the BLR material to receive the scattered ionizing radiation.}
	\label{fig:schematic}
\end{figure*}

The BLR in parametric models can have different geometries of interest depending on the purpose, for example a ring, a disk, a shell, or a cone filled with clouds in which the density of clouds is assumed to follow a certain function \citep{netzer1993, Ward2014, tek2016, Tek2018,GravityColl2018}. As the name of these models implies, the distribution, density, and other physical properties of clouds in the BLR are parameterized in order to optimize the shape of emission lines and time-delays map \citep{pancoast2011, pancoast2014, li2013}.
In order to fit the data, they either infer the transfer function \citep{Blandford1982, horne1991, krolik1995, li2016}, or provide a quantitative constraints on the dynamics and geometry of BLR \citep{li2013, pancoast2014}. 
Although these models are useful, providing us with an insight into the connection between the BLR geometry and line shape, and they can fit the observational data, they are not supposed to provide a theoretically-motivated scenario behind the picture. 

On the other hand, we previously showed that the Failed Radiatively Accelerated Dusty Outflow (FRADO) model \citep{Czerny2011}, among all other already proposed theoretical mechanisms, may effectively account for the formation and the dynamics of the LIL part of BLR such as H$\beta$ and MgII \citep{naddaf2021}. A preliminary test of the model also showed that it reproduces the reverberation-known location of the LIL BLR \citep{peterson2004, bentz2009, bentz2013} without including any arbitrary parameter, and also can successfully explain the observed dispersion in H$\beta$ radius-luminosity (RL) relation based on the accretion rate \citep{naddaf2020}. On the other hand, the high ionization lines such as CIV and HeII are located much closer to the center \citep[e.g.][]{wandel1999,grier2013} where dust can not form and dust-driving mechanism is not available so they should most likely be related to line driven winds \citep{proga2000,dyda2018}, or thermally driven winds \citep[e.g.][]{ganguly2021}.

In this paper, we aim at testing the model with calculating the line-profiles by introducing a relatively large grid table of results for a range of black hole masses, accretion rates, and dust-to-gas mass ratio of clumps. Therefore we structure the paper as follows. We briefly explain the physics of BLR dynamics in 2.5D FRADO in section \ref{sec:model}. The numerical setup is introduced in section \ref{sec:setup}. The physics of LIL BLR emission lines and the method of line profile calculation is addressed in the section \ref{sec:profiles}. We then present our results in section \ref{sec:results}, followed by discussion in section \ref{sec:discussion}.

\section{BLR Dynamics in 2.5D FRADO}\label{sec:model}

The 2.5D FRADO model \citep[see][for full details]{naddaf2021} is a nonhydrodynamical single-cloud approach to the LIL BLR dynamics.
It is the enhanced version of the basic 1D model of \cite{Czerny2011}, a model based on the existence of dust in the atmosphere of accretion disk at large radii where it is cold enough \citep{Rees1969, Dong2008}. FRADO model works on the basis of radiatively dust-driving mechanism, so the disk radiation pressure acting on the initially dusty material (clumps) at the surface layers of the accretion disk leads to a (mostly failed) outflow from the accretion disk. We should stress here that only dust processes do the job in this model, and the line-driving or electron scattering forces are negligible at these radii, hence making the model appropriate for the LIL BLR. General relativistic effect are also negligible at these large radii, so neglected.

In the 2.5D FRADO, the model is enhanced with the realistic description of the dust opacities interacting with the radiation field. Geometrical configurations as proxies for the shielding effect necessary to launch an efficient outflow, hinted by numerous authors \citep{shlosman1985,Voit1992, murray1995, proga2004, Risaliti2010, Mizumoto2019}, are incorporated into the model that can protect the dusty clumps from the intense central disk radiation, hence avoiding too rapid sublimation of the dust content of the clouds. Most AGN do not show the double peak line profiles expected from the illuminated disk surface, so in order to produce single-peak line profiles the material has to reach high vertical velocity under the continuous radiative force from below. The timescale of the process depend very much on the black hole mass as the strong radiative force must act for a period being a noticeable fraction of the local Keplerian period. Similar requirement is necessary for line-driving winds, responsible for HIL part of the BLR where shielding is needed in order to prevent over-ionization of the medium and loss of the driving power. Clumps at initially circular local Keplerian orbits are launched (with a zero vertical velocity) from the disk surface by the disk radiation pressure and move under the radiative force from the accretion disk and gravitational field of the central black hole. The radial dependent height of the disk (disk surface) for the grid of initial conditions in our model was calculated using a separate code \citep{rozanska1999, Czerny2016}.
Upon attaining high altitudes, depending on the launching radius, clumps may lose their dust content due to strong radiation from inner radii of the disk and continue their motion in the form of a free-fall in the black hole gravitational field; otherwise they remain dusty during the motion. Moreover, depending on the launching location clumps may escape into infinity or come back to disk surface but they cannot cross the disk.

The whole pattern of the trajectories of clouds depends on the initial physical parameters, and the geometry of LIL BLR in our model was determined by the kinematics of the clouds. As previously studied \citep{naddaf2021}, it does not resemble the simple (or complicated) shapes in parametric models. We showed that the overall picture of LIL BLR can be either very complex or just as simple as up/down motion depending on the accretion rate. The general pattern of motion in FRADO consists of 3 components, as indicated in figure \ref{fig:schematic}: (a) an outer tail consisting of dusty failed winds (simple up/down motions) that resembles the static puffed-up irradiated disk model of \cite{baskin2018}, this component is always present regardless of the adopted values of the initial parameters and for the very low values the full BLR is in this form; (b) an inner region with slightly to largely elongated elliptical orbits; and (c) a funnel-shaped stream of escaping clouds similar to empirical picture of AGNs \citep{elvis2000}. The last two components may develop (or not) depending on the initial parameters, i.e. the black hole mass, accretion rate, and metallicity. As previously discussed but not calculated, it can be expected that the first component may give rise to a double-peaked (disky shape) line profile consistent with observational data for low Eddington sources, however, single-peaked profiles are likely expected from the two latter components \citep[see][for more details]{naddaf2021}.

In the previous work, aimed at testing the model, the black hole mass was fixed to $10^{8} M_{\odot}$ and we studied the cases with dust-to-gas ratio equivalent to solar metalicity for a range of accretion rates. We did not investigate how changing the black hole mass and other parameters can affect the overall picture. So in this paper we present a large table of results for a relatively wide range of physical parameters of AGNs as introduced in the section \ref{sec:setup} for which we test the 2.5D FRADO model with calculation of line shapes.

\section{Numerical Setup} \label{sec:setup}

In order to have a relatively comprehensive and large table of results, the following numerical setup is considered. The main initial physical parameters in this setup, i.e. black hole mass $M_{\rm BH}$, dimensionless accretion rate $\dot m$, and dust-to-gas mass ratio $\Psi$ are: 
\begin{itemize}
    \item $M_{\rm BH}$ of $10^{6}, 10^{7}, 10^{8},$ and $10^{9} M_{\odot}$.
    \item $\dot m$ of $0.01, 0.1$ and $1$ in Eddington units defined as
    \begin{equation}
        \dot M_{\mathrm{edd}} = \dfrac{4 \pi G M_{\mathrm{BH}} m_{\rm p}}{\mu~ \sigma_{T}~ c},
    \end{equation}
    where $\mu$ is the accretion efficiency fixed to 0.1. For a reference, the Eddington value for the black hole mass of $10^{6} M_{\odot}$ is as
    \begin{equation}
        \dot M_{\mathrm{edd(6)}}= 1.399 \times 10^{24} ~~ \mathrm{[g/s]}
    \end{equation}
    \item $\Psi$ of $0.005$ and $0.025$ equivalent to 1 and 5 times solar metalicity, respectively.
\end{itemize}

The first two almost cover the observed range of black hole mass and Eddington rate of AGNs \citep{shen2011, panda2018}. On the other hand, many studies indicate that the metallicity in AGNs can super-solar \cite[see e.g.][]{hamann1992, artymowicz1993a, artymowicz1993b, matteucci1993, ferland1996, hamann1997, warner2002, dietrich2003, castro2017, xu2018, Shangguan2018,  sniegowska2020}, although an observational study by \cite{arredondo2021} shows that the value of dust-to-gas ratio for the torus can range from 0.01 to 1 times of the interstellar medium. Studies also show that there is no correlation between the metallicity in AGNs with either Eddington rate or black hole mass, or total luminosity \cite[see][and references therein]{DuPu2014}. So in this paper, in addition to solar value adopted in our previous work, we present our results for 5 times solar metallicity, as in \cite{baskin2018}.

Sublimation temperature of dust is kept at $1500$ K \citep{baskin2018} as in \citet{naddaf2021}.
The values of wavelength-dependent dust opacities kept as before are based on the prescriptions of \cite{rollig2013} and \cite{Szczerba1997} for the classical MRN dust model \citep{mathis1977} consisting of silicate and graphite grains.
The standard extended optically thick, geometrically thin disk model of \cite{SS1973} is the source of radiation in our model.
As our two previously proposed geometrical configurations for the shielding effect did not show any significant difference we arbitrarily choose the $\alpha$-patch model, with an $\alpha$ parameter equal to 3 as before for the matter of consistency \citep{naddaf2021}.

\section{LIL BLR line profiles}\label{sec:profiles}

The observed line profiles reflect the distribution of the BLR material, the velocity field and finally the local emissivity. In our model the velocity field is determined, but we still need to provide the amount of material along the trajectories, and the efficiency of the line formation at each location. Particularly this last aspect is not simple, despite many years of studies. Reverberation mapping shows that lines do respond to the irradiation by the central source, and the formation of LIL lines like H$\beta$, MgII or FeII requires incident hard continuum (far UV). However, the response of the medium to irradiation is complex. The dust radiation pressure alleviating the clouds does not guarantee any line formation in a cloud as long as the cloud does not receive some UV radiation from the central parts. Particularly in the case of LIL lines, the density of the medium is large, and in principle the collisional excitation, self-shielding, dust content, complex coupling between the continuum and line transfer, and local turbulence should be considered \citep[see e.g.][]{baldwin1997,Dragana2012}. Such a complete 3-D model is still beyond the scope of the current paper.

\subsection{Calculation of line profile}

Given a distribution of position and velocity of clouds within BLR, one can find the overall shape of line emission by the clouds. On one hand, the emission wavelengths of clouds are shifted from the rest-frame wavelength depending on their velocity component toward/against the observer due to relativistic Doppler shift. On the other hand, the position of clouds can provide some weight to the amount of central flux they receive, and also to their transparency to the observer, thereby the amount of flux re-emitted by each cloud in the form of LIL. The emission by the clouds depending on their location might be subject to gravitational redshift as well.

The Doppler shift and gravitational redshift together cause the emitted wavelength $\lambda_{\rm emit}$ of the cloud line emission to shift toward the observed wavelength $\lambda_{\rm obs}$.
However, as the radial distance of the onset of LIL BLR from the center is large, of order of several $10^2$ (in case of the largest black hole mass and the smallest accretion rate) to few $10^4$ $r_{g}$ (in case of the smallest black hole mass and the largest accretion rate), the gravitational redshift is negligible, hence, we only address the Doppler shift.

In the current paper, we assume a uniform constant cloud density within the entire LIL BLR as the line fitting in LIL BLR \citep[e.g.][]{tek2016, panda2018} supports the universal value for the cloud density. It can be further enhanced in future by assuming a certain density function for the hot surrounding medium, thereby the pressure balance can yield the density of clouds as a function of cloud position \citep{rozanska2006,baskin2018}. We also do not address the radiative transfer in our non-hydrodynamic simplified model.

The most important effect is the issue of the shielding. In the dynamics, the shielding effect is included which leads to efficient acceleration of the cloud. But fully shielded clouds do not produce emission lines. When we check a posteriori which clouds are finally well exposed to the central flux we see that only a small fraction of levitated dust-sublimated clouds in the inner LIL BLR can directly see the central UV radiation. An example of a cloud directly illuminated by the central flux is given and discussed in the section results.

Thus the illumination must be predominantly indirect. This is in agreement with other arguments that LIL part of the BLR does not see the full emission from the disk central parts. First argument comes from shorter than expected time delays in higher Eddington ratio sources and it introduces the geometrical concept of two BLR regions - one is closer to the symmetry axis and exposed to irradiation while the second one is hiding behind the geometrically thick accretion disk \citep{wang_shielding2014}. The other argument comes directly from estimates of the line equivalent width which imply that only a small fraction, of order of one percent of the nuclear emission is reprocessed to give the LIL lines like H$\beta$, FeII and CaII triplet \citep{swayam_cafeIII2021}. We thus assume that this emission comes from the scattering of the central UV radiation.

The presence of highly or fully ionized medium is well supported by observations as well as by the theory. The general picture is nicely illustrated by \citet{almeida2017} which shows the smooth outflow of low density material filling the space between clouds. This medium extends from direct vicinity of a black hole to Narrow Line region and beyond. In the case of Seyfert 2 galaxies, we see the scattering taking place in the NLR since the region closer in is still shielded from us, and the reflection reveals the presence of the BLR lines otherwise hidden from the observer \citep{antonucci1985}. In the case of type 1 sources we have a clear view down to the nucleus and we see the scatterers all the way down. Scattering medium closer in consists of hot plasma only. In studies of polarized light of Seyfert 1 galaxies these scatterers were phenomenologically divided into polar and equatorial scatterers \citep{smith2004,smith2005}. However, it well may be a continuous medium, mostly consisting of the innermost fully ionized wind plus intercloud BLR (and further away, also NLR) medium. The amount of such outflow is difficult to estimate, but arguments based on energy/momentum leads to massive outflows which can be even optically thick for electron scattering \citep{king2010}.
The polarization level is rather low, from a fraction of a percent to a few percent \citep[see e.g.][for most recent extensive measurements]{capetti2021,popovic2021}. Such a polarization level is consistent with the optical depth of scatterers or order of 1, as it largely depends on the viewing angle \citep[][]{lira2020}. From theoretical point of view, the hot plasma together with the radiation pressure provide the confinement to the clouds \citep{krolik1981,baskin2018}.

Thus the hot medium extending from the black hole vicinity up to BLR distances can scatter a fraction of the nuclear emission towards BLR clouds. Since the temperature of the medium is likely set at a Compton temperature value, $\sim 10^7$ K \citep{Rybicki1986}, only Thomson elastic scattering is important.
This radiation scattered by the surrounding hot medium reaches to the clouds as illustrated in figure \ref{fig:schematic}.

Our previous calculation of column density \citep{naddaf2021} indicates that the clouds at higher altitudes are more transparent to the scattered central flux and also to the observer. This motivates us to give some weight to the otherwise uniform intrinsic emissivity of clouds as a function of their vertical position, i.e.
\begin{equation}
    \epsilon_{\rm ~tot} = z \times \epsilon_{\rm ~int}
\end{equation}
where $z$ is the vertical position of clouds relative to the equatorial plane, and $\epsilon_{int}$ is the intrinsic constant emissivity of clouds which we assumed to be uniform throughout the LIL BLR.

As shown in figure \ref{fig:schematic}, we assume that in case of formation of stream it may block the ionizing scattered central radiation to reach to outer BLR, so no efficient line emission is expected from that part. In the absence of the stream, lines can be produced by the whole BLR.

\begin{figure*}
	\centering
	\includegraphics[scale=0.53]{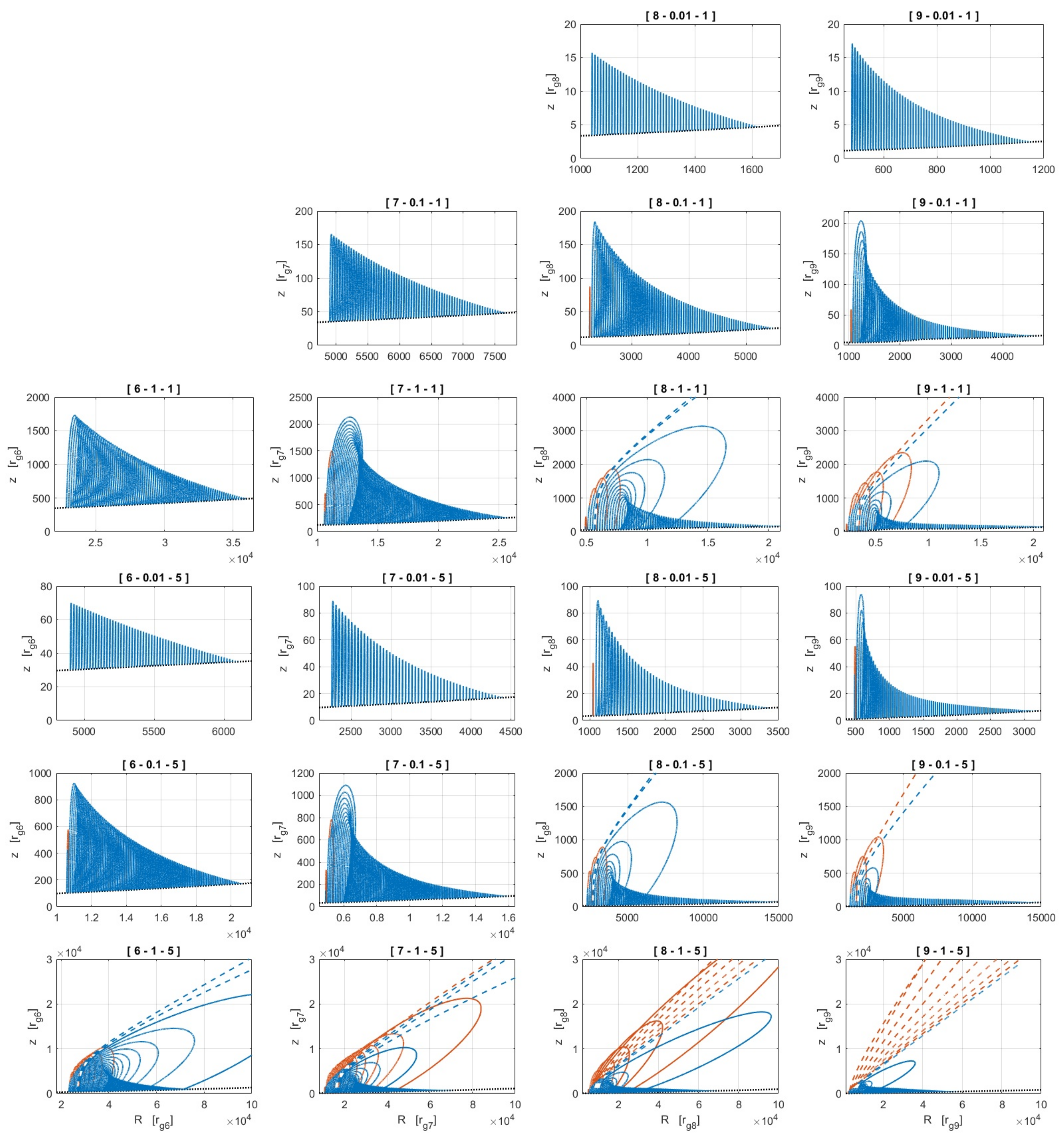}
	\caption{Trajectories of clouds within BLR in FRADO. The 3 numbers from left to right in brackets atop of each panel stand for log of black hole mass in solar units, accretion rate in Eddington units, and the metallicity in solar units, respectively. The blue and red solid/dashed lines show the path of motion of dusty and dustless failed/escaping clouds, respectively. The black dotted line represents the disk surface.}
	\label{fig:trajectories}
\end{figure*}

\begin{figure*}
	\centering
	\includegraphics[scale=0.54]{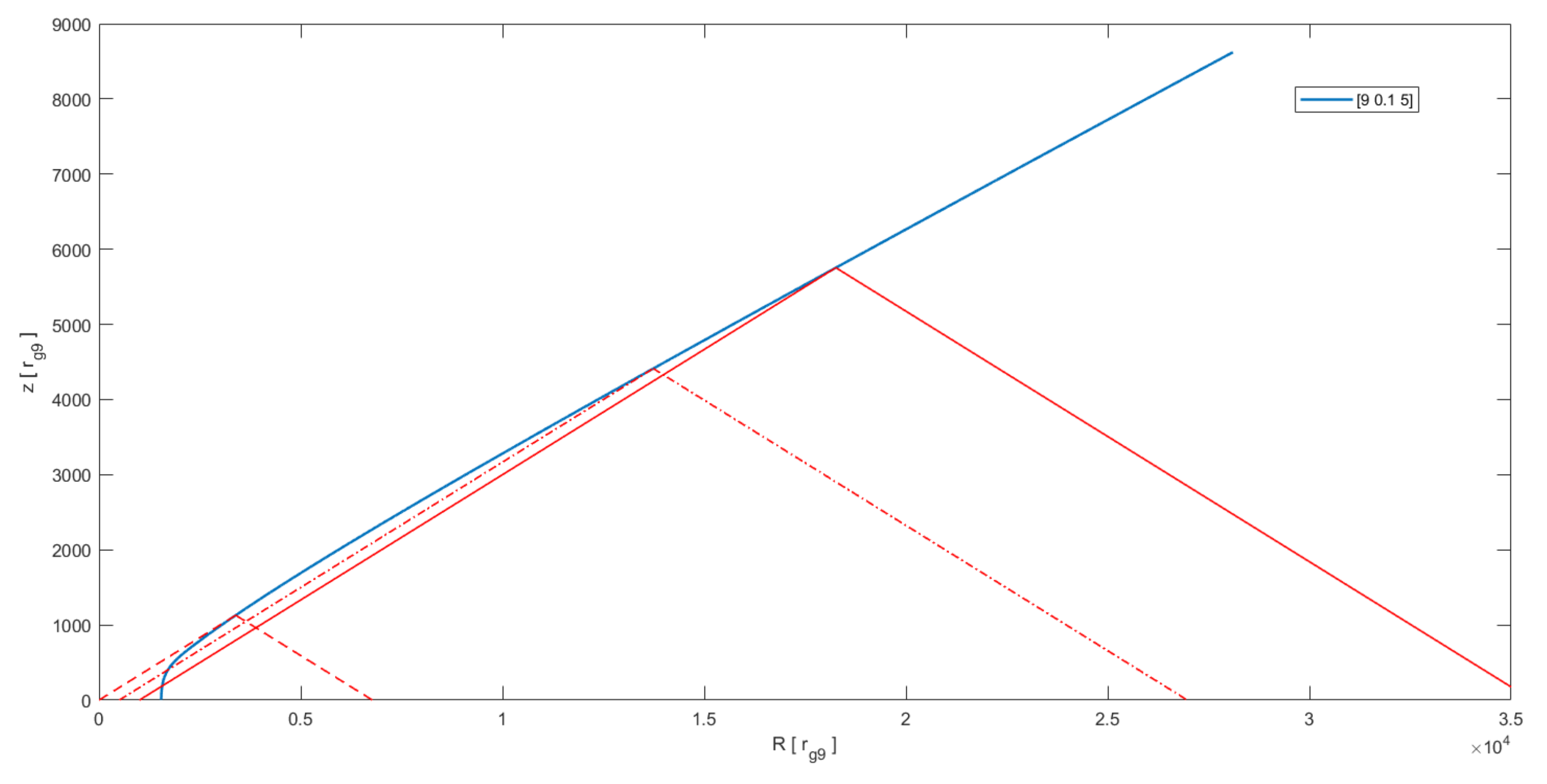}
	\caption{Illustration of the shielding effect in our model. Here we show the radial extension of the disk area seen by the selected flying cloud at 3 exemplary different positions along its trajectory, for the case of escaping cloud launched at 1540 $r_g$ with black hole mass of $10^{9} M_{\odot}$, Eddington rate of 0.1, and 5 times solar metallicity. The cloud located at vertical position of around 1100, 4500, and 5800 $r_g$ along its trajectory, sees the radial range of [0 to 6700], [500 to 26800], and [1000 to 35600] $r_g$, respectively. The model assumes the radial visibility three times larger than the local cloud height ($\alpha = 3$, see \citealt{naddaf2021}), so the area at the starting point is very small and we cannot show the early stages in this graphical scale, but the cloud is well exposed to the whole disk irradiation after reaching the height of 1000 $r_g$. Other clouds may never achieve such exposure to the disk central parts.}
	\label{fig:shield_illust}
\end{figure*}

It should be noted that in our model no line emission is expected from the disk itself due to shielding effect. Also, we do not include 'moon effect' in the line calculation. This moon effect can be the important if the individual clouds are optically thick so that 
the illuminated side of the clouds are brighter than the dark side and emits then more radiation, \citep[see e.g.][]{Goad2012, Czerny2017}. Since we have not assumed optically thick clouds in our model, all clouds emit isotropically all what they receive (absorb), in the form of line emission.
Although the model is symmetric with respect to the equatorial plane, only the clouds above the disk can contribute in the line production as the clouds on the other side are obscured by the disk itself.

\begin{figure*}
	\centering
	\includegraphics[scale=0.54]{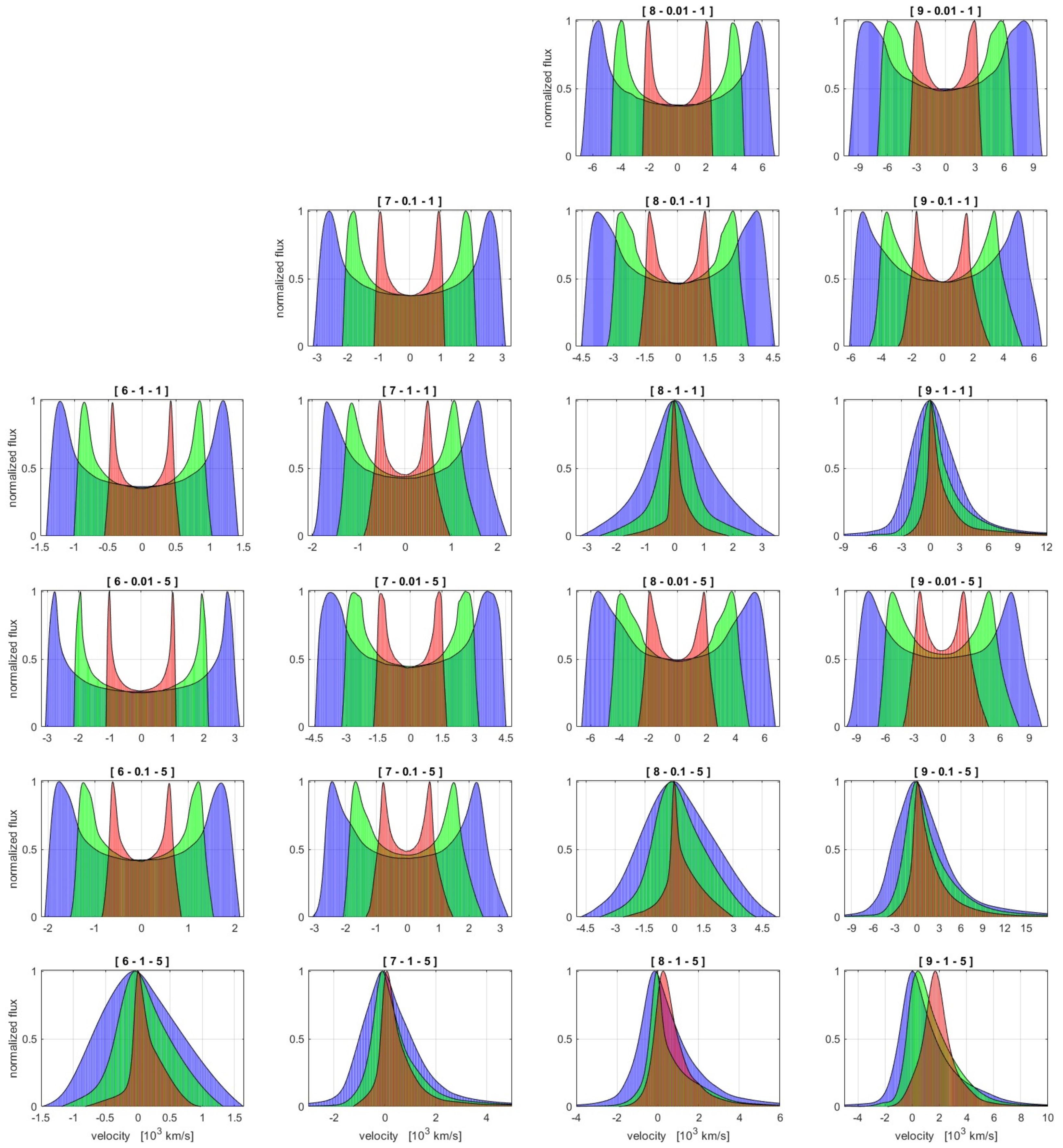}
	\caption{Predicted dependence of line Profiles (normalized flux to one) on the accretion rate, black hole mass, and dust-to-gas ratio. The numbers in brackets are as in figure \ref{fig:trajectories}. Lines are color coded with red, green, and blue for the corresponding viewing angles of 15, 30, and 45, respectively.}
	\label{fig:lineprofiles45}
\end{figure*}

\subsection{Location of clouds in FRADO}
 
In order to find the distribution of clouds in 2.5D FRADO, we need first to know the number of clouds to be launched at any given radius within the LIL BLR, or equivalently, the disk mass loss rate as a function of radius. The total mass loss rate results directly from our dynamical model \citep{naddaf2021}, however, in order to have it as a function of radius we adopt a method based on the optically-thin approximation in stellar winds. Applying this approach to accretion disk, as in \cite{Czerny2017}, yields the mass loss rate as
\begin{equation}
    \dot M_{z} \propto r^{-\frac{5}{2}}
\end{equation}

Assuming a constant cloud density, we can then calculate the total number of clouds that we have to launch per given radius in our model. We set the cloud density, arbitrarily and case-specifically, with the aim to have a total number of around two millions of clouds building the LIL BLR. Knowing the theoretical number of clouds launched at a given radius, a uniform random number generator (URNG) is then used in the following way to provide a non-biased random distribution of clouds. For each cloud launched at a radius, a URNG is used to randomly determine the azimuthal angle at which the cloud should be launched; This is due to that fact that we have azimuthal symmetry in our model but we finally need a 3D distribution of clouds since the inclined observer breaks the symmetry. We use URNG again for the same launched cloud to find its random position along its trajectory, taking into account the velocity profile along the trajectory.

\section{Results} \label{sec:results}

\subsection{Trajectories of BLR clouds}

The figure \ref{fig:trajectories} displays the trajectories of clouds launched at different radii within LIL BLR for the range of our main 3 initial physical parameters, i.e. black hole mass, accretion rate, and metallicity. 
Computation of trajectories are done for a very dense equally spaced set of initial radii within the region but only a small fraction of them are plotted for the matter of better visibility. The 3 blank panels in the figure \ref{fig:trajectories} indicate that radiation pressure was not efficient in those cases to launch any material from the disk.

The results are expressed in terms of black hole mass corresponding gravitational radius defined as
\begin{equation}
    r_{\mathrm{g}} = \dfrac{G M_{\mathrm{BH}}}{c^2}
\end{equation}
which is the minimum for the smallest black hole mass of $10^{6} M_{\odot}$ in our sample, that is
\begin{equation}
    r_{\mathrm{g(6)}} = 4.78 \times 10^{-8}~       \mathrm{[pc]} = 5.7 \times 10^{-5}~ \mathrm{[lt-day]}
\end{equation}

As can be seen, the funnel-shaped stream of material can form and get broader with an increase of not only black hole mass and accretion rate but also with an increase of the dust-to-gas ratio parameter.
The stream shows the smallest inclination (relative to the symmetry axis) for the largest values of initial parameters in our model grid, as visible in the bottom-right panel of figure \ref{fig:trajectories}. 
The radial extension of BLR, i.e. the whole radial range within which a cloud can be lifted from the disk surface, also becomes larger with the increase of the values of initial parameters. In our previous study done for the fixed black hole mass of $10^{8} M_{\odot}$, we showed that the character of motion in 2.5D FRADO strongly depends on the accretion rate. Now, for the large grid of initial parameters one can already see that it strongly depends not only on the Eddington rate of the source, but also strongly on the black hole mass and dust-to-gas ratio.

\begin{figure*}
	\centering
	\includegraphics[scale=0.6]{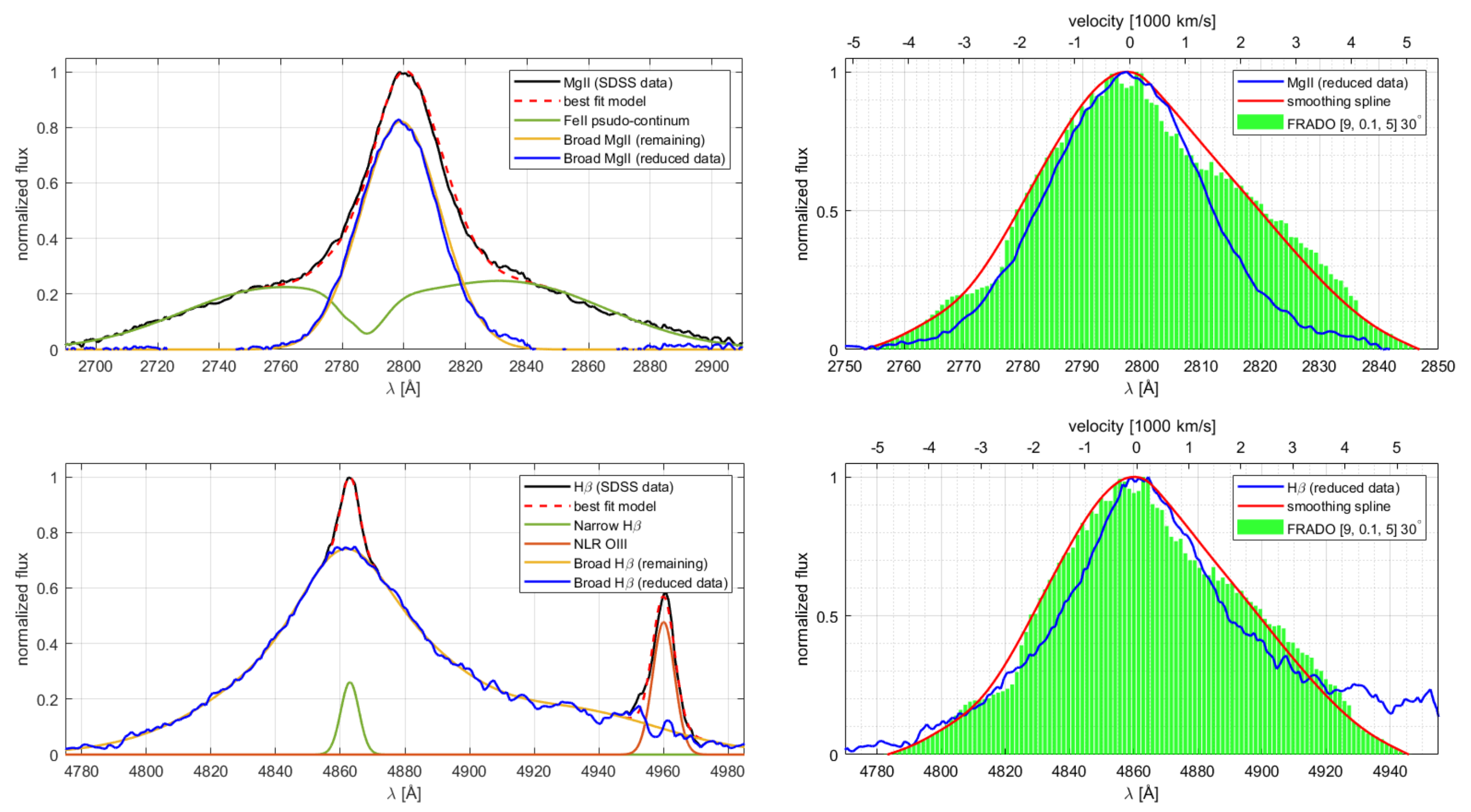}
	\caption{Right panels shows the comparison of the fully corrected MgII and H$\beta$ lines in the mean composite quasar spectrum from SDSS data to the line profile predicted by 2.5D FRADO for cases corresponding to the mean quasar physical parameters (black hole mass of $10^{9} M_{\odot}$, Eddington rate of 0.1) with metallicity of 5 times solar one, viewed at 30 degrees. In the left panels, the power-law-subtracted MgII and H$\beta$ lines from SDSS data are decomposed into different components as indicated in the legends of plots.}
	\label{fig:composite_spect}
\end{figure*}

As seen from the results plotted in figure \ref{fig:trajectories}, there are 8 cases in our model grid, for different set of initial parameters, for which the component c (i.e. the outflow stream) is formed so that the component a (i.e. the outer tail) is not expected to contribute in line production as it does not receive the ionizing scattered radiation.

We here also provide a scaled illustration of the action of shielding effect in detail in figure \ref{fig:shield_illust} for the case of an escaping trajectory for a model with black hole mass of $10^{9} M_{\odot}$, accretion rate of 0.1 in Eddington units, and 5 times solar metalicity. We draw there the increasing fraction of the disk visible to the cloud as it moves along the trajectory. This cloud, when reaching the height of around $10^3 r_g$ is illuminated by the central disk flux. Generally in our model, the clouds for high mass, high accretion rate and high metallicity can/may be fully directly irradiated when they reach relatively large distances from the center. But low mass, low Eddington ratio, low metallicity solutions never predict strong direct exposure of the clouds to full irradiation; In such cases, from the observational point of view, LIL lines do form as well and show the double-peak profiles suggesting the line origin to be close to the disk surface while direct irradiation in this case is highly inefficient (see e.g. \citealt{loska2004}). This is consistent with the findings from our model.

\subsection{Line profiles}

\begin{figure}
	\centering
	\includegraphics[scale=0.53]{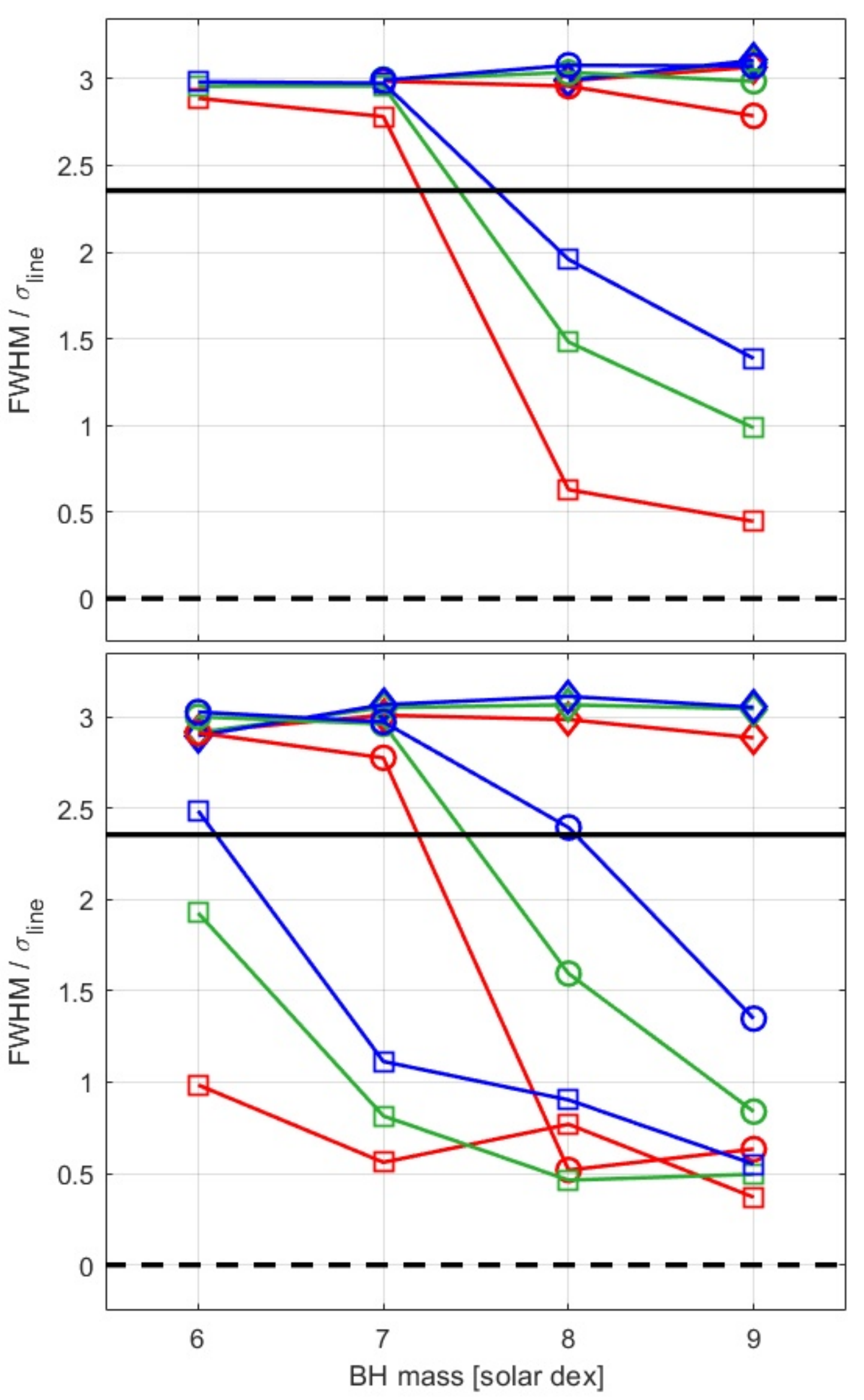}
	\caption{Dependence of the ratio of $\rm FWHM$/$\sigma_{\rm line}$ on the initial physical parameters for the solar (upper panel) and 5 times solar metallicity (lower panel). The red, green, and blue solid lines represents the viewing angles of 15, 30, and 45 degrees, respectively. The diamonds, circles, and squares stand for the Eddington rates of 0.01, 0.1, and 1, respectively. The values of 2.3548 and zero expected for a Gaussian and Lorentzian profile are depicted by black solid and black dashed lines, respectively, for a reference. }
	\label{fig:FWHM}
\end{figure}

The figure \ref{fig:lineprofiles45} shows the full grid of predicted line profiles for the distribution of clumps building the LIL BLR seen at different viewing angles. The line shapes shown in this figure were smoothed with spline technique for better appearance. 

We have set the values of 15, 30, and 45 degree as the representative values of viewing angle for the type 1 AGNs. For these sources the viewing angle is never very large as the BLR is then obscured by the torus \citep{antonucci1993, netzer2015}.

Consistently with observation, the lines get narrower with Eddington ratio \citep{pounds1995, Dupu2016} of the source and become broader with black hole mass. Except for the high Eddington rate and high black hole masses, the predicted lines at solar metallicity are dominated by the disk-like shape. However, the higher value of metallicity changes the picture as it allows the efficient rise of material above the disk with vertical velocities comparable to rotational Keplerian one.
Generally, an overall view shows that all cases in which the component c of BLR (as shown in figure \ref{fig:schematic}) is developed, a single-peaked profile can be expected; other cases without this component give rise to double-peaked line profiles.
Comparing the line shapes by eye for the metallicity, one can argue that the increase of the metallicity by a factor of 5 corresponds to the increase of the Eddington ratio of the source by 10 times. Likewise, one can expect single-peaked profiles for all super-Eddington sources regardless of the black hole mass and metallicity.

There are two important features in the single-peaked predicted profiles: asymmetry and blue-shift.
The asymmetry can not be due to the difference between the time duration of rise and fall of the clouds since in all cases only a very small fraction of failed clouds pass the sublimation location and lose their dust content and continue their motion as a ballistic motion. It means most of clouds finish their full orbit keeping their dust content so that their motion are not expected to cause line asymmetry. As can be implied from both figures \ref{fig:trajectories} and \ref{fig:lineprofiles45}, the more intense and broader the outflow stream is the more asymmetry and blue-shift in the line shape is visible. So these two features are interrelated and they imply that the outflow is dominant compared to failed part of BLR.

\section{Discussion} \label{sec:discussion}

\subsection{Comparison with 1D FRADO}

The line shapes predicted by 2.5D FRADO model show considerable improvement approaching to characteristic Lorentzian profiles for high Eddington high black hole mass sources compared to 1D FRADO in which the double-peaked structure was always yielded \citep{Czerny2017}.

Compared to 1D model with constant wavelength-averaged opacities, the 2.5D model is enhanced with the advanced realistic wavelength-dependent values of dust opacities where the dust-to-gas ratio regulates the strength of the radiative force. Increasing the dust-to-gas ratio from solar value to 5 times solar (appropriate for quasars as discussed before) leads to a major improvement of the line shape toward single-peaked profiles characteristic for high accretors; whereas this could be obtained in 1D model if the wavelength-averaged dust opacity was hugely increased by a (not likely realistic) factor of 1000 \citep{Czerny2017}.

\subsection{Composite Spectrum (Mean Quasar)}

The mean composite quasar spectrum \citep{compositespectra2001} from the Sloan Digital Sky Survey (SDSS) shows that a single-peaked line profile is expected for the LIL BLR. Such a spectrum well represents quasars in SDSS data \citep{shen2011}, where the mean value of the black hole mass is $10^9 M_{\odot}$ (Swayamtrupta Panda, private communication), and the mean Eddington ratio is 0.1 \citep{panda2018}.
We thus compared the model taking the parameters appropriate for the mean quasar to the MgII and H$\beta$ line shape, as shown in figure \ref{fig:composite_spect}. With this aim we locally subtracted the underlying power-law, and then over-plotted the theoretical shape of the line assuming arbitrary normalization.
In the case of H$\beta$ line, prior removal of the narrow H$\beta$ component as well as [OIII] lines coming from the NLR was necessary, and for that we fixed the width of these components at 205 km/s and 220 km/s for Hbeta and OIII, respectively.
As for the MgII line, we used the semi-empirical UV FeII template\footnote{\url{http://servo.aob.rs/FeII_AGN/link7.html}} \citep{kovacevic2015, popovic2019} to remove the FeII pseudo-continuum in the required spectral band. We see that the results show a nice fit with the data for the LIL part of BLR.

\subsection{$\rm FWHM$ vs. $\sigma_{\rm line}$}

The ratio of the $\rm FWHM$ to the line dispersion $\sigma_{\rm line}$ is an important parameter \citep{collin2006} characterizing the line profile, although it does not differentiate between a single and double-peak profiles. We calculated this ratio for the grid of our initial physical parameters. The figure \ref{fig:FWHM} shows the results. Most models are roughly consistent with the ratio expected for a Gaussian, but for high mass high Eddington ratio models this ratio dropped considerable, indicating that we approach the Lorentzian profiles.
Therefore, although the Lorentzian profiles might be basically only expected for extremely low viewing angles \citep[e.g.][]{Goad2012}, our results show that sources with high mass high Eddington ratio can generally lead to relatively Lorentzian shapes, regardless of the viewing angle. Moreover, the larger the black hole mass and accretion rate are, the more asymmetry and blue-shift in the line shape we have, as shown in figure \ref{fig:lineprofiles45}

This trend is overall consistent with the data \citep[see e.g.][]{marziani2003}. The strong trend with the change of the viewing angle is visible for high Eddington sources while for the remaining sources the effect is weak, when $\rm FWHM$/$\sigma_{\rm line}$ is used. Somewhat surprising is the very strong trend with the black hole mass which is also visible in figure~\ref{fig:lineprofiles45}. 

\subsection{Model assumptions}

In our model we use non-hydrodynamical approach based on assumption of the motion of separate clouds under gravity and the action of radiation pressure acting on dust. This has considerable limitations but they are justified as the first approximation for modelling the LIL part of the BLR. As well discussed in the classical paper of line driven wind model (for HIL part of BLR) of \citet{murray1995}, the optical depth of the emitting region must be moderate (column density of order of $10^{23}$ cm$^{-2}$), the local density is high (for LIL part it is higher than that of HIL, many authors argue for the local density about $10^{12}$ cm$^{-3}$, e.g. \citealt{adhikari2016, baskin2018, panda2018}) while the BLR is extended. There are two possibilities to create a consistent picture: either to assume of a very narrow stream of material flowing out, with the cross-section of order of $10^{12}$ cm, as in \citet{murray1995} (there they assume lower density so the size is actually larger $10^{14}$ cm), or to assume considerable clumpiness of the medium. We follow the second approach since there are natural thermal instabilities in the plasma, like instability caused by X-ray irradiation \citep{krolik1981}. In this case the plasma spontaneously forms colder clumps (at temperature $\sim 10^4$ K, cooled through atomic processes) embedded in a hotter medium (at temperature $\sim 10^7$ K, set at Inverse Compton temperature value). The presence of highly or fully ionized medium is well supported by observations as well as by the theory. Then the two media of density contrast of order of $10^3$ provide the rough pressure balance. \citet{blandford1990} discusses the typical values for ionization parameter in AGN clouds covering the range from $10^{-3}$ to 1, i.e. up to four orders of magnitudes.The precise description of the structure of the clumpy medium is very difficult. Even in the case of a single cloud exposed to irradiation, in plane-parallel approximation requires performing the radiative transfer which shows then the gradual change of the density, temperature and ionization parameter \citep[see e.g.][]{baskin2018, Tek2018}, with the low density first, and the temperature roughly at Inverse Compton temperature (which depends on the shape of the incident spectrum), then relatively rapid decrease at the subsequent ionization fronts. The proper description of this transition, calculated under constant pressure, actually requires inclusion of the electron conduction \citep[e.g.][]{begelman1990,rozanska2000}. Deep inside the cloud there is some further drop in the temperature and a rise in the density caused by the decrease of the local flux due to absorption. As stressed by \citet{baskin2018}, also the radiation pressure plays a dynamical role. The picture further complicates if plane-parallel approximation is abandoned. The presence of the numerous clouds of complex shape can be fully consistent with simple estimates of the cloud number based on line shape properties as done by \citet{arav1998}. Of course, there are also processes leading to cloud destruction, e.g. the action of tidal forces \citep{muller2022}, Kelvin-Helmholtz instabilities, and cloud ablation) but the destruction rate can be strongly affected by the magnetic field \citep[e.g.][]{mccourt2015}. The relative importance between the condensation rate and destruction rate depends on the cloud size, as it is set roughly by the Field length \citep{field1965}. Cloud formation in AGN was seen in numerical simulations of \citet{waters2021}, but at distances much larger than the BLR distance which was likely related to the numerical setup and the requested spatial resolution of the computations. The issue is thus extremely complex, simple order of magnitude estimates based on a single density and single temperature of the cloud and intercloud medium are not fully adequate and cannot reproduce the full ionization parameter range, but addressing this point in detail is beyond the scope of the current paper. 

Our model has one more important simplification. We assume only the action of radiation pressure acting on the dust, and we neglect the line driving effect. Close to the launching point the local flux comes from the disk, without considerable UV component, but when the clouds is alleviated high above the disk, the line-driving effect should set in. There were numerous papers addressing the line-driven winds (\citealt{murray1995, proga2000, Risaliti2010}; see \citealt{Giustini2021} for a recent review) but none of the papers combined dust radiation pressure and line pressure in this context since the way of determining the radiative force is very different in these two scenarios. Our negligence of the line driving is most likely appropriate for description of launching clouds at LIL BLR but it underestimated the cloud velocities when the outflowing stream of the material forms. This effect should be incorporated in the future studies (specifically, we should combine our code with the QWIND \citep{Risaliti2010, qwind3} but this is not easy even in the case of non-hydro simulations.

Expected higher velocities of the clouds due to line driving will partially be slowed down by the interaction of the clouds with the ambient medium, which is also neglected in our model. The freshly launched clouds and inter-cloud medium are co-moving since clouds form through thermal instability. However, accelerating clouds detach from their surrounding, and thereby some drag force is expected, although clouds are much denser than the ambient medium, roughly in pressure equilibrium, and the relative velocities between the hot medium and the clouds are not very high since clouds and inter-cloud medium likely share the Keplerian motion component of the velocity and follow the same initial vertical velocity gained before the fragmentation of the medium into clouds occurred.

Wind models having elements of the radiative transfer show single-peaked lines \citep{murray1995, waters2016}. In our work, with a simplified non-hydrodynamic approach to the dynamics of material in LIL BLR, the single-peaked profiles are obtained without incorporating the process of radiative transfer into the emissivity of clouds.

In the future work it will be necessary to address the issue of cloud emissivity in more detail, including the line formation in the dusty medium. However, this effect is strongly dependent not only on cloud irradiation but also on cloud local densities. At low densities, the presence of the dust suppresses the line emission \citep{netzer1993} while at high densities (above $10^{11.5}$ cm$^{-3}$) emission lines do form efficiently even in the presence of the dust which explains no gap between the NLR and BLR region in Narrow Line Seyfert 1 galaxies \citep{adhikari2016}.

The assumption of the dust sublimation temperature fixed in the at 1500 K is simple to relax, and higher/lower temperature will predominantly shift the position of the BLR outward/inward. This will be important for precise comparison of the radius-luminosity relation to the observational data, but we do not yet focus here on this goal. Indeed, higher average temperature of $\sim 1700$ K is quite likely, but 1500 K is still frequently assumed even in data analysis \citep[e.g.][]{dexter_GRAVITY2020}, and broad band spectral fitting of the dust emission in AGN gives the values in individual sources in the range of 1000 - 1750 K \citep{hernan2016}. The more advanced approach, taking into account the specific dust species, like silicates, amorphous carbon etc, and the dependence of the sublimation temperature not only on the chemical composition but also on specific grain size is far more complicated since the loss of the driving pressure force would be gradual, and the dust composition would have to be recalculated at each integration step for each cloud trajectory. This should be done at some stage, but a number of simplifications still used to recover the dynamics and the line emissivity do not request yet such advanced approach.

\section{Summary}

Following our previous work on the dynamical character and shape of LIL BLR based on the non-hydrodynamic single-cloud 2.5D FRADO model \citep{naddaf2021}, we here tested the model by calculation of line profiles for the model-concluded distribution of clouds along their trajectories within LIL BLR, for a relatively large grid of initial conditions. We adopted a simple approach with constant cloud density and without the element of radiative transfer included.

The predicted trend in the line shape seems consistent with observations, as the model implies narrower/broader line shape for sources with larger Eddington ratio/smaller black hole mass. All solutions with a developed outflow stream of material show up single-peaked profiles, and the line shape approaches a Lorentzian shape for high mass high Eddington ratio. In other cases double-peaked disk-like line shape dominates.

Two features: asymmetry and blue-shift seen in the line shapes developed in the cases with outflow streams show that the outflow is dominant compared to outflow/inflow (failed trajectories). These features were more visibly intense for the sources with high black hole mass high Eddington ratios. 

Most importantly, we showed that the line profile predicted by our model for the case with initial physical parameters corresponding to those of the mean quasar is consistent with observed mean spectrum seen in the SDSS composite.

\begin{acknowledgements}
      The project was partially supported by the Polish Funding Agency National Science Centre, project 2017/26/A/ST9/00756 (MAESTRO 9), and MNiSW grant DIR/WK/2018/12. Authors would like to thank Swayamtrupta Panda for his help and fruitful discussion on quasar populations.
\end{acknowledgements}

\bibliographystyle{aa}
\bibliography{naddaf}

\begin{thebibliography}{119}
\expandafter\ifx\csname natexlab\endcsname\relax\def\natexlab#1{#1}\fi

\bibitem[{{Adhikari} {et~al.}(2018){Adhikari}, {Hryniewicz},
  {R{\'o}{\.z}a{\'n}ska}, {Czerny}, \& {Ferland}}]{Tek2018}
{Adhikari}, T.~P., {Hryniewicz}, K., {R{\'o}{\.z}a{\'n}ska}, A., {Czerny}, B.,
  \& {Ferland}, G.~J. 2018, \apj, 856, 78

\bibitem[{{Adhikari} {et~al.}(2016{\natexlab{a}}){Adhikari},
  {R{\'o}{\.z}a{\'n}ska}, {Czerny}, {Hryniewicz}, \& {Ferland}}]{tek2016}
{Adhikari}, T.~P., {R{\'o}{\.z}a{\'n}ska}, A., {Czerny}, B., {Hryniewicz}, K.,
  \& {Ferland}, G.~J. 2016{\natexlab{a}}, \apj, 831, 68

\bibitem[{{Adhikari} {et~al.}(2016{\natexlab{b}}){Adhikari},
  {R{\'o}{\.z}a{\'n}ska}, {Czerny}, {Hryniewicz}, \& {Ferland}}]{adhikari2016}
{Adhikari}, T.~P., {R{\'o}{\.z}a{\'n}ska}, A., {Czerny}, B., {Hryniewicz}, K.,
  \& {Ferland}, G.~J. 2016{\natexlab{b}}, \apj, 831, 68

\bibitem[{{Antonucci}(1993)}]{antonucci1993}
{Antonucci}, R. 1993, ARA\&A, 31, 473

\bibitem[{{Antonucci} \& {Miller}(1985)}]{antonucci1985}
{Antonucci}, R.~R.~J. \& {Miller}, J.~S. 1985, \apj, 297, 621

\bibitem[{{Arav} {et~al.}(1998){Arav}, {Barlow}, {Laor}, {Sargent}, \&
  {Blandford}}]{arav1998}
{Arav}, N., {Barlow}, T.~A., {Laor}, A., {Sargent}, W. L.~W., \& {Blandford},
  R.~D. 1998, \mnras, 297, 990

\bibitem[{{Artymowicz}(1993)}]{artymowicz1993b}
{Artymowicz}, P. 1993, \pasp, 105, 1032

\bibitem[{{Artymowicz} {et~al.}(1993){Artymowicz}, {Lin}, \&
  {Wampler}}]{artymowicz1993a}
{Artymowicz}, P., {Lin}, D.~N.~C., \& {Wampler}, E.~J. 1993, \apj, 409, 592

\bibitem[{{Baldwin}(1997)}]{baldwin1997}
{Baldwin}, J.~A. 1997, in Astronomical Society of the Pacific Conference
  Series, Vol. 113, IAU Colloq. 159: Emission Lines in Active Galaxies: New
  Methods and Techniques, ed. B.~M. {Peterson}, F.-Z. {Cheng}, \& A.~S.
  {Wilson}, 80

\bibitem[{{Baskin} \& {Laor}(2018)}]{baskin2018}
{Baskin}, A. \& {Laor}, A. 2018, MNRAS, 474, 1970

\bibitem[{{Begelman} \& {McKee}(1990)}]{begelman1990}
{Begelman}, M.~C. \& {McKee}, C.~F. 1990, \apj, 358, 375

\bibitem[{{Bentz} {et~al.}(2013){Bentz}, {Denney}, {Grier}, {Barth},
  {Peterson}, {Vestergaard}, {Bennert}, {Canalizo}, {De Rosa}, {Filippenko},
  {Gates}, {Greene}, {Li}, {Malkan}, {Pogge}, {Stern}, {Treu}, \&
  {Woo}}]{bentz2013}
{Bentz}, M.~C., {Denney}, K.~D., {Grier}, C.~J., {et~al.} 2013, \apj, 767, 149

\bibitem[{{Bentz} {et~al.}(2009){Bentz}, {Walsh}, {Barth}, {Baliber},
  {Bennert}, {Canalizo}, {Filippenko}, {Ganeshalingam}, {Gates}, {Greene},
  {Hidas}, {Hiner}, {Lee}, {Li}, {Malkan}, {Minezaki}, {Sakata}, {Serduke},
  {Silverman}, {Steele}, {Stern}, {Street}, {Thornton}, {Treu}, {Wang}, {Woo},
  \& {Yoshii}}]{bentz2009}
{Bentz}, M.~C., {Walsh}, J.~L., {Barth}, A.~J., {et~al.} 2009, ApJ, 705, 199

\bibitem[{{Blandford} \& {McKee}(1982)}]{Blandford1982}
{Blandford}, R.~D. \& {McKee}, C.~F. 1982, \apj, 255, 419

\bibitem[{{Blandford} {et~al.}(1990){Blandford}, {Netzer}, {Woltjer},
  {Courvoisier}, \& {Mayor}}]{blandford1990}
{Blandford}, R.~D., {Netzer}, H., {Woltjer}, L., {Courvoisier}, T.~J.~L., \&
  {Mayor}, M. 1990, in Active Galactic Nuclei

\bibitem[{{Boroson} \& {Green}(1992)}]{Boroson1992}
{Boroson}, T.~A. \& {Green}, R.~F. 1992, \apjs, 80, 109

\bibitem[{{Capetti} {et~al.}(2021){Capetti}, {Laor}, {Baldi}, {Robinson}, \&
  {Marconi}}]{capetti2021}
{Capetti}, A., {Laor}, A., {Baldi}, R.~D., {Robinson}, A., \& {Marconi}, A.
  2021, \mnras, 502, 5086

\bibitem[{{Castro} {et~al.}(2017){Castro}, {Dors}, {Cardaci}, \&
  {H{\"a}gele}}]{castro2017}
{Castro}, C.~S., {Dors}, O.~L., {Cardaci}, M.~V., \& {H{\"a}gele}, G.~F. 2017,
  \mnras, 467, 1507

\bibitem[{{Collin} {et~al.}(2006){Collin}, {Kawaguchi}, {Peterson}, \&
  {Vestergaard}}]{collin2006}
{Collin}, S., {Kawaguchi}, T., {Peterson}, B.~M., \& {Vestergaard}, M. 2006,
  \aap, 456, 75

\bibitem[{{Collin-Souffrin} {et~al.}(1988){Collin-Souffrin}, {Dyson},
  {McDowell}, \& {Perry}}]{CollinSouffrin1988}
{Collin-Souffrin}, S., {Dyson}, J.~E., {McDowell}, J.~C., \& {Perry}, J.~J.
  1988, MNRAS, 232, 539

\bibitem[{{Czerny} {et~al.}(2016){Czerny}, {Du}, {Wang}, \&
  {Karas}}]{Czerny2016}
{Czerny}, B., {Du}, P., {Wang}, J.-M., \& {Karas}, V. 2016, ApJ, 832, 15

\bibitem[{{Czerny} \& {Hryniewicz}(2011)}]{Czerny2011}
{Czerny}, B. \& {Hryniewicz}, K. 2011, \aap, 525, L8

\bibitem[{{Czerny} {et~al.}(2017){Czerny}, {Li}, {Hryniewicz}, {Panda},
  {Wildy}, {Sniegowska}, {Wang}, {Sredzinska}, \& {Karas}}]{Czerny2017}
{Czerny}, B., {Li}, Y.-R., {Hryniewicz}, K., {et~al.} 2017, ApJ, 846, 154

\bibitem[{{Dietrich} {et~al.}(2003){Dietrich}, {Hamann}, {Shields},
  {Constantin}, {Heidt}, {J{\"a}ger}, {Vestergaard}, \&
  {Wagner}}]{dietrich2003}
{Dietrich}, M., {Hamann}, F., {Shields}, J.~C., {et~al.} 2003, \apj, 589, 722

\bibitem[{{Dong} {et~al.}(2008){Dong}, {Wang}, {Wang}, {Yuan}, {Zhou}, {Dai},
  \& {Zhang}}]{Dong2008}
{Dong}, X., {Wang}, T., {Wang}, J., {et~al.} 2008, MNRAS, 383, 581

\bibitem[{{Du} {et~al.}(2016){Du}, {Lu}, {Zhang}, {Huang}, {Wang}, {Hu}, {Qiu},
  {Li}, {Fan}, {Fang}, {Bai}, {Bian}, {Yuan}, {Ho}, {Wang}, \& {SEAMBH
  Collaboration}}]{Dupu2016}
{Du}, P., {Lu}, K.-X., {Zhang}, Z.-X., {et~al.} 2016, ApJ, 825, 126

\bibitem[{{Du} {et~al.}(2014){Du}, {Wang}, {Hu}, {Valls-Gabaud}, {Baldwin},
  {Ge}, \& {Xue}}]{DuPu2014}
{Du}, P., {Wang}, J.-M., {Hu}, C., {et~al.} 2014, \mnras, 438, 2828

\bibitem[{{Dyda} \& {Proga}(2018)}]{dyda2018}
{Dyda}, S. \& {Proga}, D. 2018, \mnras, 475, 3786

\bibitem[{{Elvis}(2000)}]{elvis2000}
{Elvis}, M. 2000, \apj, 545, 63

\bibitem[{{Esparza Arredondo} {et~al.}(2021){Esparza Arredondo}, {Gonz{\'a}lez
  Mart{\'\i}n}, {Dultzin}, {Masegosa}, {Ramos Almeida}, {Garc{\'\i}a Bernete},
  {Fritz}, \& {Osorio Clavijo}}]{arredondo2021}
{Esparza Arredondo}, D., {Gonz{\'a}lez Mart{\'\i}n}, O., {Dultzin}, D.,
  {et~al.} 2021, arXiv e-prints, arXiv:2104.11263

\bibitem[{{Ferland} {et~al.}(1996){Ferland}, {Baldwin}, {Korista}, {Hamann},
  {Carswell}, {Phillips}, {Wilkes}, \& {Williams}}]{ferland1996}
{Ferland}, G.~J., {Baldwin}, J.~A., {Korista}, K.~T., {et~al.} 1996, \apj, 461,
  683

\bibitem[{{Field}(1965)}]{field1965}
{Field}, G.~B. 1965, \apj, 142, 531

\bibitem[{{Ganguly} {et~al.}(2021){Ganguly}, {Proga}, {Waters}, {Dannen},
  {Dyda}, {Giustini}, {Kallman}, {Raymond}, {Miller}, \& {Rodriguez
  Hidalgo}}]{ganguly2021}
{Ganguly}, S., {Proga}, D., {Waters}, T., {et~al.} 2021, \apj, 914, 114

\bibitem[{{Gaskell}(2009)}]{Gaskell2009}
{Gaskell}, C.~M. 2009, \nar, 53, 140

\bibitem[{{gezari} {et~al.}(2007){gezari}, {Halpern}, \&
  {Eracleous}}]{gezari2007}
{gezari}, S., {Halpern}, J.~P., \& {Eracleous}, M. 2007, \apjs, 169, 167

\bibitem[{{Giustini} \& {Proga}(2021)}]{Giustini2021}
{Giustini}, M. \& {Proga}, D. 2021, in Nuclear Activity in Galaxies Across
  Cosmic Time, ed. M.~{Povi{\'c}}, P.~{Marziani}, J.~{Masegosa}, H.~{Netzer},
  S.~H. {Negu}, \& S.~B. {Tessema}, Vol. 356, 82--86

\bibitem[{{Goad} {et~al.}(2012){Goad}, {Korista}, \& {Ruff}}]{Goad2012}
{Goad}, M.~R., {Korista}, K.~T., \& {Ruff}, A.~J. 2012, \mnras, 426, 3086

\bibitem[{{Gravity Collaboration} {et~al.}(2021){Gravity Collaboration},
  {Amorim}, {Baub{\"o}ck}, {Brandner}, {Bolzer}, {Cl{\'e}net}, {Davies}, {de
  Zeeuw}, {Dexter}, {Drescher}, {Eckart}, {Eisenhauer}, {F{\"o}rster
  Schreiber}, {Gao}, {Garcia}, {Genzel}, {Gillessen}, {Gratadour}, {H{\"o}nig},
  {Kaltenbrunner}, {Kishimoto}, {Lacour}, {Lutz}, {Millour}, {Netzer}, {Ott},
  {Paumard}, {Perraut}, {Perrin}, {Peterson}, {Petrucci}, {Pfuhl}, {Prieto},
  {Rouan}, {Sanchez-Bermudez}, {Shangguan}, {Shimizu}, {Schartmann}, {Stadler},
  {Sternberg}, {Straub}, {Straubmeier}, {Sturm}, {Tacconi}, {Tristram},
  {Vermot}, {von Fellenberg}, {Waisberg}, {Widmann}, \&
  {Woillez}}]{GRAVITY3783_2021}
{Gravity Collaboration}, {Amorim}, A., {Baub{\"o}ck}, M., {et~al.} 2021, \aap,
  648, A117

\bibitem[{{Gravity Collaboration} {et~al.}(2020{\natexlab{a}}){Gravity
  Collaboration}, {Amorim}, {Baub{\"o}ck}, {Brandner}, {Cl{\'e}net}, {Davies},
  {de Zeeuw}, {Dexter}, {Eckart}, {Eisenhauer}, {F{\"o}rster Schreiber}, {Gao},
  {Garcia}, {Genzel}, {Gillessen}, {Gratadour}, {H{\"o}nig}, {Kishimoto},
  {Lacour}, {Lutz}, {Millour}, {Netzer}, {Ott}, {Paumard}, {Perraut}, {Perrin},
  {Peterson}, {Petrucci}, {Pfuhl}, {Prieto}, {Rouan}, {Shangguan}, {Shimizu},
  {Schartmann}, {Stadler}, {Sternberg}, {Straub}, {Straubmeier}, {Sturm},
  {Tacconi}, {Tristram}, {Vermot}, {von Fellenberg}, {Waisberg}, {Widmann}, \&
  {Woillez}}]{GRAVITYIRAS_2020}
{Gravity Collaboration}, {Amorim}, A., {Baub{\"o}ck}, M., {et~al.}
  2020{\natexlab{a}}, \aap, 643, A154

\bibitem[{{Gravity Collaboration} {et~al.}(2020{\natexlab{b}}){Gravity
  Collaboration}, {Dexter}, {Shangguan}, {H{\"o}nig}, {Kishimoto}, {Lutz},
  {Netzer}, {Davies}, {Sturm}, {Pfuhl}, {Amorim}, {Baub{\"o}ck}, {Brandner},
  {Cl{\'e}net}, {de Zeeuw}, {Eckart}, {Eisenhauer}, {F{\"o}rster Schreiber},
  {Gao}, {Garcia}, {Genzel}, {Gillessen}, {Gratadour}, {Jim{\'e}nez-Rosales},
  {Lacour}, {Millour}, {Ott}, {Paumard}, {Perraut}, {Perrin}, {Peterson},
  {Petrucci}, {Prieto}, {Rouan}, {Schartmann}, {Shimizu}, {Sternberg},
  {Straub}, {Straubmeier}, {Tacconi}, {Tristram}, {Vermot}, {Waisberg},
  {Widmann}, \& {Woillez}}]{dexter_GRAVITY2020}
{Gravity Collaboration}, {Dexter}, J., {Shangguan}, J., {et~al.}
  2020{\natexlab{b}}, \aap, 635, A92

\bibitem[{{Gravity Collaboration} {et~al.}(2018{\natexlab{a}}){Gravity
  Collaboration}, {Sturm}, {Dexter}, {Pfuhl}, {Stock}, {Davies}, {Lutz},
  {Cl{\'e}net}, {Eckart}, {Eisenhauer}, {Genzel}, {Gratadour}, {H{\"o}nig},
  {Kishimoto}, {Lacour}, {Millour}, {Netzer}, {Perrin}, {Peterson}, {Petrucci},
  {Rouan}, {Waisberg}, {Woillez}, {Amorim}, {Brandner}, {F{\"o}rster
  Schreiber}, {Garcia}, {Gillessen}, {Ott}, {Paumard}, {Perraut},
  {Scheithauer}, {Straubmeier}, {Tacconi}, \& {Widmann}}]{GRAVITY3C273_2018}
{Gravity Collaboration}, {Sturm}, E., {Dexter}, J., {et~al.}
  2018{\natexlab{a}}, \nat, 563, 657

\bibitem[{{Gravity Collaboration} {et~al.}(2018{\natexlab{b}}){Gravity
  Collaboration}, {Sturm}, {Dexter}, {Pfuhl}, {Stock}, {Davies}, {Lutz},
  {Cl{\'e}net}, {Eckart}, {Eisenhauer}, {Genzel}, {Gratadour}, {H{\"o}nig},
  {Kishimoto}, {Lacour}, {Millour}, {Netzer}, {Perrin}, {Peterson}, {Petrucci},
  {Rouan}, {Waisberg}, {Woillez}, {Amorim}, {Brandner}, {F{\"o}rster
  Schreiber}, {Garcia}, {Gillessen}, {Ott}, {Paumard}, {Perraut},
  {Scheithauer}, {Straubmeier}, {Tacconi}, \& {Widmann}}]{GravityColl2018}
{Gravity Collaboration}, {Sturm}, E., {Dexter}, J., {et~al.}
  2018{\natexlab{b}}, \nat, 563, 657

\bibitem[{{Grier} {et~al.}(2013){Grier}, {Peterson}, {Horne}, {Bentz}, {Pogge},
  {Denney}, {De Rosa}, {Martini}, {Kochanek}, {Zu}, {Shappee}, {Siverd},
  {Beatty}, {Sergeev}, {Kaspi}, {Araya Salvo}, {Bird}, {Bord}, {Borman}, {Che},
  {Chen}, {Cohen}, {Dietrich}, {Doroshenko}, {Efimov}, {Free}, {Ginsburg},
  {Henderson}, {King}, {Mogren}, {Molina}, {Mosquera}, {Nazarov}, {Okhmat},
  {Pejcha}, {Rafter}, {Shields}, {Skowron}, {Szczygiel}, {Valluri}, \& {van
  Saders}}]{grier2013}
{Grier}, C.~J., {Peterson}, B.~M., {Horne}, K., {et~al.} 2013, ApJ, 764, 47

\bibitem[{{Guerras} {et~al.}(2013){Guerras}, {Mediavilla}, {Jimenez-Vicente},
  {Kochanek}, {Mu{\~n}oz}, {Falco}, \& {Motta}}]{guerras2013}
{Guerras}, E., {Mediavilla}, E., {Jimenez-Vicente}, J., {et~al.} 2013, \apj,
  764, 160

\bibitem[{{Hamann}(1997)}]{hamann1997}
{Hamann}, F. 1997, \apjs, 109, 279

\bibitem[{{Hamann} \& {Ferland}(1992)}]{hamann1992}
{Hamann}, F. \& {Ferland}, G. 1992, \apjl, 391, L53

\bibitem[{{Hern{\'a}n-Caballero} {et~al.}(2016){Hern{\'a}n-Caballero},
  {Hatziminaoglou}, {Alonso-Herrero}, \& {Mateos}}]{hernan2016}
{Hern{\'a}n-Caballero}, A., {Hatziminaoglou}, E., {Alonso-Herrero}, A., \&
  {Mateos}, S. 2016, \mnras, 463, 2064

\bibitem[{{Horne} {et~al.}(1991){Horne}, {Welsh}, \& {Peterson}}]{horne1991}
{Horne}, K., {Welsh}, W.~F., \& {Peterson}, B.~M. 1991, \apjl, 367, L5

\bibitem[{{Ili{\'c}} {et~al.}(2012){Ili{\'c}}, {Popovi{\'c}}, {La Mura},
  {Ciroi}, \& {Rafanelli}}]{Dragana2012}
{Ili{\'c}}, D., {Popovi{\'c}}, L.~{\v{C}}., {La Mura}, G., {Ciroi}, S., \&
  {Rafanelli}, P. 2012, \aap, 543, A142

\bibitem[{{Kaspi} {et~al.}(2000){Kaspi}, {Smith}, {Netzer}, {Maoz}, {Jannuzi},
  \& {Giveon}}]{kaspi2000}
{Kaspi}, S., {Smith}, P.~S., {Netzer}, H., {et~al.} 2000, \apj, 533, 631

\bibitem[{{King}(2010)}]{king2010}
{King}, A.~R. 2010, \mnras, 402, 1516

\bibitem[{{Kova{\v{c}}evi{\'c}-Doj{\v{c}}inovi{\'c}} \&
  {Popovi{\'c}}(2015)}]{kovacevic2015}
{Kova{\v{c}}evi{\'c}-Doj{\v{c}}inovi{\'c}}, J. \& {Popovi{\'c}}, L.~{\v{C}}.
  2015, \apjs, 221, 35

\bibitem[{{Krolik} \& {Done}(1995)}]{krolik1995}
{Krolik}, J.~H. \& {Done}, C. 1995, \apj, 440, 166

\bibitem[{{Krolik} {et~al.}(1981){Krolik}, {McKee}, \& {Tarter}}]{krolik1981}
{Krolik}, J.~H., {McKee}, C.~F., \& {Tarter}, C.~B. 1981, \apj, 249, 422

\bibitem[{{Lawrence} {et~al.}(1997){Lawrence}, {Elvis}, {Wilkes}, {McHardy}, \&
  {Brandt}}]{Lawrence1997}
{Lawrence}, A., {Elvis}, M., {Wilkes}, B.~J., {McHardy}, I., \& {Brandt}, N.
  1997, \mnras, 285, 879

\bibitem[{{Le} \& {Woo}(2019)}]{Le2019}
{Le}, H. A.~N. \& {Woo}, J.-H. 2019, \apj, 887, 236

\bibitem[{{Li} {et~al.}(2021){Li}, {Yang}, {Yang}, {Chen}, {Songsheng}, {Liu},
  {Du}, {Luo}, {Yu}, {Hu}, {Jiang}, {Bao}, {Guo}, {Zhang}, {Li}, {Xiao}, {Lu},
  {Ho}, {Bai}, {Bian}, {Aceituno}, {Minezaki}, {Horne}, {Kokubo}, \&
  {Wang}}]{li2021}
{Li}, S.-S., {Yang}, S., {Yang}, Z.-X., {et~al.} 2021, \apj, 920, 9

\bibitem[{{Li} {et~al.}(2016){Li}, {Wang}, \& {Bai}}]{li2016}
{Li}, Y.-R., {Wang}, J.-M., \& {Bai}, J.-M. 2016, \apj, 831, 206

\bibitem[{{Li} {et~al.}(2013){Li}, {Wang}, {Ho}, {Du}, \& {Bai}}]{li2013}
{Li}, Y.-R., {Wang}, J.-M., {Ho}, L.~C., {Du}, P., \& {Bai}, J.-M. 2013, \apj,
  779, 110

\bibitem[{{Lira} {et~al.}(2020){Lira}, {Goosmann}, {Kishimoto}, \&
  {Cartier}}]{lira2020}
{Lira}, P., {Goosmann}, R.~W., {Kishimoto}, M., \& {Cartier}, R. 2020, \mnras,
  491, 1

\bibitem[{{Loska} {et~al.}(2004){Loska}, {Czerny}, \& {Szczerba}}]{loska2004}
{Loska}, Z., {Czerny}, B., \& {Szczerba}, R. 2004, \mnras, 355, 1080

\bibitem[{{Lu} {et~al.}(2021){Lu}, {Wang}, {Zhang}, {Huang}, {Xu}, {Xin}, {Yu},
  {Ding}, {Wang}, \& {Feng}}]{lu2021}
{Lu}, K.-X., {Wang}, J.-G., {Zhang}, Z.-X., {et~al.} 2021, \apj, 918, 50

\bibitem[{{Marziani} {et~al.}(2003){Marziani}, {Sulentic}, {Zamanov},
  {Calvani}, {Dultzin-Hacyan}, {Bachev}, \& {Zwitter}}]{marziani2003}
{Marziani}, P., {Sulentic}, J.~W., {Zamanov}, R., {et~al.} 2003, \apjs, 145,
  199

\bibitem[{{Mathis} {et~al.}(1977){Mathis}, {Rumpl}, \&
  {Nordsieck}}]{mathis1977}
{Mathis}, J.~S., {Rumpl}, W., \& {Nordsieck}, K.~H. 1977, \apj, 217, 425

\bibitem[{{Matteucci} \& {Padovani}(1993)}]{matteucci1993}
{Matteucci}, F. \& {Padovani}, P. 1993, \apj, 419, 485

\bibitem[{{McCourt} {et~al.}(2015){McCourt}, {O'Leary}, {Madigan}, \&
  {Quataert}}]{mccourt2015}
{McCourt}, M., {O'Leary}, R.~M., {Madigan}, A.-M., \& {Quataert}, E. 2015,
  \mnras, 449, 2

\bibitem[{{Mizumoto} {et~al.}(2019){Mizumoto}, {Done}, {Tomaru}, \&
  {Edwards}}]{Mizumoto2019}
{Mizumoto}, M., {Done}, C., {Tomaru}, R., \& {Edwards}, I. 2019, \mnras, 489,
  1152

\bibitem[{{M{\"u}ller} {et~al.}(2022){M{\"u}ller}, {Naddaf}, {Zaja{\v{c}}ek},
  {Czerny}, {Araudo}, \& {Karas}}]{muller2022}
{M{\"u}ller}, A.~L., {Naddaf}, M.-H., {Zaja{\v{c}}ek}, M., {et~al.} 2022, arXiv
  e-prints, arXiv:2204.05361

\bibitem[{{Murray} {et~al.}(1995){Murray}, {Chiang}, {Grossman}, \&
  {Voit}}]{murray1995}
{Murray}, N., {Chiang}, J., {Grossman}, S.~A., \& {Voit}, G.~M. 1995, \apj,
  451, 498

\bibitem[{{Naddaf} {et~al.}(2020){Naddaf}, {Czerny}, \&
  {Szczerba}}]{naddaf2020}
{Naddaf}, M.-H., {Czerny}, B., \& {Szczerba}, R. 2020, Frontiers in Astronomy
  and Space Sciences, 7, 15

\bibitem[{{Naddaf} {et~al.}(2021){Naddaf}, {Czerny}, \&
  {Szczerba}}]{naddaf2021}
{Naddaf}, M.-H., {Czerny}, B., \& {Szczerba}, R. 2021, \apj, 920, 30

\bibitem[{{Negrete} {et~al.}(2018){Negrete}, {Dultzin}, {Marziani}, {Esparza},
  {Sulentic}, {del Olmo}, {Mart{\'\i}nez-Aldama}, {Garc{\'\i}a L{\'o}pez},
  {D'Onofrio}, {Bon}, \& {Bon}}]{negrete2018}
{Negrete}, C.~A., {Dultzin}, D., {Marziani}, P., {et~al.} 2018, \aap, 620, A118

\bibitem[{{Netzer}(2013)}]{netzer2013}
{Netzer}, H. 2013, {The Physics and Evolution of Active Galactic Nuclei}
  (Cambridge, UK: Cambridge University Press)

\bibitem[{{Netzer}(2015)}]{netzer2015}
{Netzer}, H. 2015, ARA\&A, 53, 365

\bibitem[{{Netzer}(2020)}]{netzer2020}
{Netzer}, H. 2020, \mnras, 494, 1611

\bibitem[{{Netzer} \& {Laor}(1993)}]{netzer1993}
{Netzer}, H. \& {Laor}, A. 1993, \apjl, 404, L51

\bibitem[{{Osterbrock}(1977)}]{osterbrock1977}
{Osterbrock}, D.~E. 1977, ApJ, 215, 733

\bibitem[{{Osterbrock}(1981)}]{osterbrock1981}
{Osterbrock}, D.~E. 1981, ApJ, 249, 462

\bibitem[{{Pancoast} {et~al.}(2011){Pancoast}, {Brewer}, \&
  {Treu}}]{pancoast2011}
{Pancoast}, A., {Brewer}, B.~J., \& {Treu}, T. 2011, \apj, 730, 139

\bibitem[{{Pancoast} {et~al.}(2014){Pancoast}, {Brewer}, \&
  {Treu}}]{pancoast2014}
{Pancoast}, A., {Brewer}, B.~J., \& {Treu}, T. 2014, \mnras, 445, 3055

\bibitem[{{Panda}(2021)}]{swayam_cafeIII2021}
{Panda}, S. 2021, \aap, 650, A154

\bibitem[{{Panda} {et~al.}(2018){Panda}, {Czerny}, {Adhikari}, {Hryniewicz},
  {Wildy}, {Kuraszkiewicz}, \& {{\'S}niegowska}}]{panda2018}
{Panda}, S., {Czerny}, B., {Adhikari}, T.~P., {et~al.} 2018, \apj, 866, 115

\bibitem[{{Peterson} {et~al.}(2004){Peterson}, {Ferrarese}, {Gilbert}, {Kaspi},
  {Malkan}, {Maoz}, {Merritt}, {Netzer}, {Onken}, {Pogge}, {Vestergaard}, \&
  {Wandel}}]{peterson2004}
{Peterson}, B.~M., {Ferrarese}, L., {Gilbert}, K.~M., {et~al.} 2004, ApJ, 613,
  682

\bibitem[{{Popovi{\'c}} {et~al.}(2019){Popovi{\'c}},
  {Kova{\v{c}}evi{\'c}-Doj{\v{c}}inovi{\'c}}, \&
  {Mar{\v{c}}eta-Mandi{\'c}}}]{popovic2019}
{Popovi{\'c}}, L.~{\v{C}}., {Kova{\v{c}}evi{\'c}-Doj{\v{c}}inovi{\'c}}, J., \&
  {Mar{\v{c}}eta-Mandi{\'c}}, S. 2019, \mnras, 484, 3180

\bibitem[{{Popovi{\'c}} {et~al.}(2021){Popovi{\'c}}, {Shablovinskaya}, \&
  {Savi{\'c}}}]{popovic2021}
{Popovi{\'c}}, L.~C., {Shablovinskaya}, E., \& {Savi{\'c}}, D. 2021, arXiv
  e-prints, arXiv:2111.06237

\bibitem[{{Pounds} {et~al.}(1995){Pounds}, {Done}, \& {Osborne}}]{pounds1995}
{Pounds}, K.~A., {Done}, C., \& {Osborne}, J.~P. 1995, \mnras, 277, L5

\bibitem[{{Proga} \& {Kallman}(2004)}]{proga2004}
{Proga}, D. \& {Kallman}, T.~R. 2004, \apj, 616, 688

\bibitem[{{Proga} {et~al.}(2000){Proga}, {Stone}, \& {Kallman}}]{proga2000}
{Proga}, D., {Stone}, J.~M., \& {Kallman}, T.~R. 2000, \apj, 543, 686

\bibitem[{{Quera-Bofarull} {et~al.}(2021){Quera-Bofarull}, {Done}, {Lacey},
  {Nomura}, \& {Ohsuga}}]{qwind3}
{Quera-Bofarull}, A., {Done}, C., {Lacey}, C.~G., {Nomura}, M., \& {Ohsuga}, K.
  2021, arXiv e-prints, arXiv:2111.02742

\bibitem[{{Raimundo} {et~al.}(2020){Raimundo}, {Vestergaard}, {Goad}, {Grier},
  {Williams}, {Peterson}, \& {Treu}}]{Raimundo2020}
{Raimundo}, S.~I., {Vestergaard}, M., {Goad}, M.~R., {et~al.} 2020, \mnras,
  493, 1227

\bibitem[{{Ramos Almeida} \& {Ricci}(2017)}]{almeida2017}
{Ramos Almeida}, C. \& {Ricci}, C. 2017, Nature Astronomy, 1, 679

\bibitem[{{Rees} {et~al.}(1969){Rees}, {Silk}, {Werner}, \&
  {Wickramasinghe}}]{Rees1969}
{Rees}, M.~J., {Silk}, J.~I., {Werner}, M.~W., \& {Wickramasinghe}, N.~C. 1969,
  \nat, 223, 788

\bibitem[{{Reeves} \& {Turner}(2000)}]{Reeves2000}
{Reeves}, J.~N. \& {Turner}, M.~J.~L. 2000, \mnras, 316, 234

\bibitem[{{Risaliti} \& {Elvis}(2010)}]{Risaliti2010}
{Risaliti}, G. \& {Elvis}, M. 2010, A\&A, 516, A89

\bibitem[{{R{\"o}llig} {et~al.}(2013){R{\"o}llig}, {Szczerba}, {Ossenkopf}, \&
  {Gl{\"u}ck}}]{rollig2013}
{R{\"o}llig}, M., {Szczerba}, R., {Ossenkopf}, V., \& {Gl{\"u}ck}, C. 2013,
  \aap, 549, A85

\bibitem[{{R{\'o}{\.z}a{\'n}ska} \& {Czerny}(2000)}]{rozanska2000}
{R{\'o}{\.z}a{\'n}ska}, A. \& {Czerny}, B. 2000, \mnras, 316, 473

\bibitem[{{R{\'o}{\.z}a{\'n}ska} {et~al.}(1999){R{\'o}{\.z}a{\'n}ska},
  {Czerny}, {{\.Z}ycki}, \& {Pojma{\'n}ski}}]{rozanska1999}
{R{\'o}{\.z}a{\'n}ska}, A., {Czerny}, B., {{\.Z}ycki}, P.~T., \&
  {Pojma{\'n}ski}, G. 1999, \mnras, 305, 481

\bibitem[{{R{\'o}{\.z}a{\'n}ska} {et~al.}(2006){R{\'o}{\.z}a{\'n}ska},
  {Goosmann}, {Dumont}, \& {Czerny}}]{rozanska2006}
{R{\'o}{\.z}a{\'n}ska}, A., {Goosmann}, R., {Dumont}, A.~M., \& {Czerny}, B.
  2006, \aap, 452, 1

\bibitem[{{Rybicki} \& {Lightman}(1986)}]{Rybicki1986}
{Rybicki}, G.~B. \& {Lightman}, A.~P. 1986, {Radiative Processes in
  Astrophysics} (Wiley-VCH)

\bibitem[{{Shakura} \& {Sunyaev}(1973)}]{SS1973}
{Shakura}, N.~I. \& {Sunyaev}, R.~A. 1973, A\&A, 500, 33

\bibitem[{{Shangguan} {et~al.}(2018){Shangguan}, {Ho}, \&
  {Xie}}]{Shangguan2018}
{Shangguan}, J., {Ho}, L.~C., \& {Xie}, Y. 2018, ApJ, 854, 158

\bibitem[{{Shen} {et~al.}(2011){Shen}, {Richards}, {Strauss}, {Hall},
  {Schneider}, {Snedden}, {Bizyaev}, {Brewington}, {Malanushenko},
  {Malanushenko}, {Oravetz}, {Pan}, \& {Simmons}}]{shen2011}
{Shen}, Y., {Richards}, G.~T., {Strauss}, M.~A., {et~al.} 2011, ApJs, 194, 45

\bibitem[{{Shlosman} {et~al.}(1985){Shlosman}, {Vitello}, \&
  {Shaviv}}]{shlosman1985}
{Shlosman}, I., {Vitello}, P.~A., \& {Shaviv}, G. 1985, \apj, 294, 96

\bibitem[{{Sluse} {et~al.}(2012){Sluse}, {Hutsem{\'e}kers}, {Courbin},
  {Meylan}, \& {Wambsganss}}]{sluse2012}
{Sluse}, D., {Hutsem{\'e}kers}, D., {Courbin}, F., {Meylan}, G., \&
  {Wambsganss}, J. 2012, \aap, 544, A62

\bibitem[{{Smith} {et~al.}(2004){Smith}, {Robinson}, {Alexander}, {Young},
  {Axon}, \& {Corbett}}]{smith2004}
{Smith}, J.~E., {Robinson}, A., {Alexander}, D.~M., {et~al.} 2004, \mnras, 350,
  140

\bibitem[{{Smith} {et~al.}(2005){Smith}, {Robinson}, {Young}, {Axon}, \&
  {Corbett}}]{smith2005}
{Smith}, J.~E., {Robinson}, A., {Young}, S., {Axon}, D.~J., \& {Corbett}, E.~A.
  2005, \mnras, 359, 846

\bibitem[{{{\'S}niegowska} {et~al.}(2020){{\'S}niegowska}, {Marziani},
  {Czerny}, {Panda}, {Mart{\'\i}nez-Aldama}, {del Olmo}, \&
  {D'Onofrio}}]{sniegowska2020}
{{\'S}niegowska}, M., {Marziani}, P., {Czerny}, B., {et~al.} 2020, arXiv
  e-prints, arXiv:2009.14177

\bibitem[{{Sulentic} {et~al.}(2000){Sulentic}, {Marziani}, \&
  {Dultzin-Hacyan}}]{Sulentic2000}
{Sulentic}, J.~W., {Marziani}, P., \& {Dultzin-Hacyan}, D. 2000, \araa, 38, 521

\bibitem[{{Szczerba} {et~al.}(1997){Szczerba}, {Omont}, {Volk}, {Cox}, \&
  {Kwok}}]{Szczerba1997}
{Szczerba}, R., {Omont}, A., {Volk}, K., {Cox}, P., \& {Kwok}, S. 1997, A\&A,
  317, 859

\bibitem[{{Vanden Berk} {et~al.}(2001){Vanden Berk}, {Richards}, {Bauer},
  {Strauss}, {Schneider}, {Heckman}, {York}, {Hall}, {Fan}, {Knapp},
  {Anderson}, {Annis}, {Bahcall}, {Bernardi}, {Briggs}, {Brinkmann}, {Brunner},
  {Burles}, {Carey}, {Castander}, {Connolly}, {Crocker}, {Csabai}, {Doi},
  {Finkbeiner}, {Friedman}, {Frieman}, {Fukugita}, {Gunn}, {Hennessy},
  {Ivezi{\'c}}, {Kent}, {Kunszt}, {Lamb}, {Leger}, {Long}, {Loveday}, {Lupton},
  {Meiksin}, {Merelli}, {Munn}, {Newberg}, {Newcomb}, {Nichol}, {Owen}, {Pier},
  {Pope}, {Rockosi}, {Schlegel}, {Siegmund}, {Smee}, {Snir}, {Stoughton},
  {Stubbs}, {SubbaRao}, {Szalay}, {Szokoly}, {Tremonti}, {Uomoto}, {Waddell},
  {Yanny}, \& {Zheng}}]{compositespectra2001}
{Vanden Berk}, D.~E., {Richards}, G.~T., {Bauer}, A., {et~al.} 2001, \aj, 122,
  549

\bibitem[{{Voit}(1992)}]{Voit1992}
{Voit}, G.~M. 1992, \mnras, 258, 841

\bibitem[{{Wandel} {et~al.}(1999){Wandel}, {Peterson}, \&
  {Malkan}}]{wandel1999}
{Wandel}, A., {Peterson}, B.~M., \& {Malkan}, M.~A. 1999, \apj, 526, 579

\bibitem[{{Wang} {et~al.}(2014){Wang}, {Qiu}, {Du}, \&
  {Ho}}]{wang_shielding2014}
{Wang}, J.-M., {Qiu}, J., {Du}, P., \& {Ho}, L.~C. 2014, \apj, 797, 65

\bibitem[{{Ward} {et~al.}(2014){Ward}, {Wadsley}, \& {Sills}}]{Ward2014}
{Ward}, R.~L., {Wadsley}, J., \& {Sills}, A. 2014, \mnras, 445, 1575

\bibitem[{{Warner} {et~al.}(2002){Warner}, {Hamann}, {Shields}, {Constantin},
  {Foltz}, \& {Chaffee}}]{warner2002}
{Warner}, C., {Hamann}, F., {Shields}, J.~C., {et~al.} 2002, \apj, 567, 68

\bibitem[{{Waters} {et~al.}(2016){Waters}, {Kashi}, {Proga}, {Eracleous},
  {Barth}, \& {Greene}}]{waters2016}
{Waters}, T., {Kashi}, A., {Proga}, D., {et~al.} 2016, \apj, 827, 53

\bibitem[{{Waters} {et~al.}(2021){Waters}, {Proga}, \& {Dannen}}]{waters2021}
{Waters}, T., {Proga}, D., \& {Dannen}, R. 2021, \apj, 914, 62

\bibitem[{{Xu} {et~al.}(2018){Xu}, {Bian}, {Shen}, {Zuo}, {Fan}, \&
  {Zhu}}]{xu2018}
{Xu}, F., {Bian}, F., {Shen}, Y., {et~al.} 2018, \mnras, 480, 345

\bibitem[{{Zhang} {et~al.}(2019){Zhang}, {Du}, {Smith}, {Zhao}, {Hu}, {Xiao},
  {Li}, {Huang}, {Wang}, {Bai}, {Ho}, \& {Wang}}]{zhang2019}
{Zhang}, Z.-X., {Du}, P., {Smith}, P.~S., {et~al.} 2019, \apj, 876, 49

\end{thebibliography}

\end{document}